%% file: main.tex
\documentclass[lettersize,journal]{IEEEtran}

\usepackage{xspace}
\usepackage{soul}
\usepackage[normalem]{ulem}
\usepackage{amsmath}
\usepackage{tikz}
\usepackage{tikz}
\usepackage{booktabs}
\usepackage{multirow}
\usepackage{enumitem}
\usepackage[hidelinks]{hyperref}

\definecolor{steelblue}{HTML}{4682B4}
\definecolor{linkColor}{HTML}{257E98}
\definecolor{quoteColor}{HTML}{C46299}

\input{sections/_preemble}

\graphicspath{{figs}{figs/old/}{figures/}{pictures/}{images/}{./}}

\usepackage{amsmath,amsfonts}
\usepackage{algorithmic}
\usepackage{array}

\usepackage{textcomp}
\usepackage{stfloats}
\usepackage{url}
\usepackage{verbatim}
\usepackage{graphicx}
\def\BibTeX{{\rm B\kern-.05em{\sc i\kern-.025em b}\kern-.08em
    T\kern-.1667em\lower.7ex\hbox{E}\kern-.125emX}}
\usepackage{balance}

\newcommand{\toolname}{FocalLens}

\begin{document}

\title{FocalLens: An Interactive Visual Interface for Understanding Narrative Perspectives}

\maketitle

\begin{abstract}
Existing visualization tools for narrative analysis often focus on interactions between characters and locations over a timeline. However, a narrative is more than characters and locations and includes complex components such as perspectives, causality, and discourse. Visualizing these components can lead to new insights that are not available in existing tools. To explore this hypothesis, in this paper, we aim to visualize a narrative through focalization, the component that captures perspective changes  (who sees or perceives the events) in a narrative. Based on a formative study with four literary scholars, we developed \toolname{}, an interactive visual interface powered by large-language models (LLMs) that shows how a story can be told from different types of perspectives (e.g., through a character's internal senses and thoughts vs. indirect actions). In a user study with 13 literary scholars, we compared \toolname{}  with Story Ribbons, an LLM-powered existing tool for visualizing and analyzing narratives. Participants found a large number of  insights about perspective analysis and writing style using \toolname{}. In contrast, participants found more insights about plot structure and characters while using Story Ribbons, although \toolname{} also facilitated some of these insights.  

\end{abstract}


\input{sections/01-introduction}
\input{sections/02-background}
\input{sections/03-related-work}
\input{sections/04-design-of-focallens}
\input{sections/05-implementation}
\input{sections/06-evaluation}
\input{sections/07-discussion}
\input{sections/08-conclusion}

\bibliographystyle{IEEEtran}
\bibliography{main}

\end{document}

%% file: sections/_preemble.tex
\newcommand{\figref}[1]{\hyperref[#1]{Fig.~\ref*{#1}}}
\newcommand{\secref}[1]{\hyperref[#1]{Sec.~\ref*{#1}}}

\usepackage[normalem]{ulem}
\usepackage{amsmath}

\newcommand{\fix}[1]{\textcolor{black}{#1}}
\newcommand{\fixx}[1]{\textcolor{black}{#1}}

\usepackage{stfloats}

\usepackage[caption=false,font=footnotesize]{subfig}

\NewDocumentCommand{\ourhref}{s m m}{%
  \href{#2}{\textcolor{steelblue}{%
    \IfBooleanT{#1}{[}#3~$\nearrow$\IfBooleanT{#1}{]}%
  }}%
}

\usepackage[table]{xcolor}
\usepackage{xcolor}
\usepackage{booktabs}
\usepackage{tabularx}

\definecolor{barA}{HTML}{3B6FA0}   % FocalLens (muted blue)
\definecolor{barB}{HTML}{E69F00}   % StoryRibbons (warm orange)
\definecolor{barnum}{HTML}{555555} % count labels

\newlength{\barunit}
\newcommand{\countbar}[2]{%
  \textcolor{#2}{\rule[0.15ex]{#1\barunit}{1.6mm}}%
  \hspace{1mm}{\textcolor{barnum}{#1}}%
}
\newcommand{\counts}[2]{%
  \begin{tabular}[t]{@{}l@{}}
    \countbar{#1}{barA}\\[-1pt]
    \countbar{#2}{orange}
  \end{tabular}%
}

\newcommand{\swatch}[1]{\textcolor{#1}{\rule[0.1ex]{2.2mm}{1.6mm}}}

\newcommand{\exrows}[2]{%
  \hangindent=3.2mm\hangafter=1
  \makebox[3.2mm][l]{\swatch{barA}}#1\par\vspace{1pt}%
  \hangindent=3.2mm\hangafter=1
  \makebox[3.2mm][l]{\swatch{orange}}#2%
}

\author{S M Raihanul Alam,
        Md Dilshadur Rahman,
        Enamul Hoque Prince,
        and Naimul Hoque%
\thanks{S M Raihanul Alam and Naimul Hoque are with the University of Iowa, Iowa City, IA, USA (e-mail: \{smraalam, nhoque\}@uiowa.edu).}%
\thanks{Md Dilshadur Rahman is with Harvard Medical School, Boston, MA, USA (e-mail: ddilshadur@gmail.com).}%
\thanks{Enamul Hoque Prince is with York University, Toronto, ON, Canada (e-mail: enamulh@yorku.ca).}%
}

%% file: sections/01-introduction.tex
\section{Introduction}
\label{sec:intro}

A \textit{story} is a sequence of events, while a \textit{narrative} is the way the story is told to the audience~\cite{rimmonkenan2002narrativefiction}. There is a linear order between the events in a story, but the narrative can break that order and reorganize the events to improve the engagement of the story. One way to understand a narrative is to decompose it into constituent components such as time, characters, locations, events,  emotion, point of view, focalization, and causality~\cite{rimmonkenan2002narrativefiction, bal2009narratology, prince2003dictionarynarratology}. It is the interplay among these components that brings stories to life and engages audiences.

Visualizing story and narrative has received significant attention in the visualization community. The most notable representation in this area is the \textit{Storyline} visualization~\cite{yeh2025story, DBLP:journals/tvcg/TanahashiM12, DBLP:conf/gd/GronemannJLM16, dibartolomeo2026survey}, which represents a narrative as a set of lines that move horizontally across time, where each line corresponds to a character. When characters interact in a scene, their lines converge and run close together. When they are not interacting, their lines separate. There exist several extensions to the Storyline visualization. For example, Story Ribbons~\cite{yeh2025story} is the LLM-based implementation of Storyline visualizations. Story Curves~\cite{DBLP:journals/tvcg/KimBISGP18} visualizes both story and narrative order. StoryPrint~\cite{DBLP:conf/iui/WatsonSSGMK19} visualizes character presence and emotion across scenes along a circular timeline, while Portrayal~\cite{hoque2023portrayal} uses heatmaps and word clouds of character attributes to represent narratives. A common feature of these representations is the use of characters and plot locations over a timeline, with some variations, to depict the narrative structure.

While these representations support analysis of character and plot dynamics, other narrative components, such as perspective, discourse, and causality~\cite{rimmonkenan2002narrativefiction}, remain largely unexplored in these solutions. The visual representation of these components can facilitate new insights about hidden patterns and linguistic styles that may otherwise remain implicit. 

\begin{figure*}[t]
    \centering
    \includegraphics[width=0.9\linewidth]{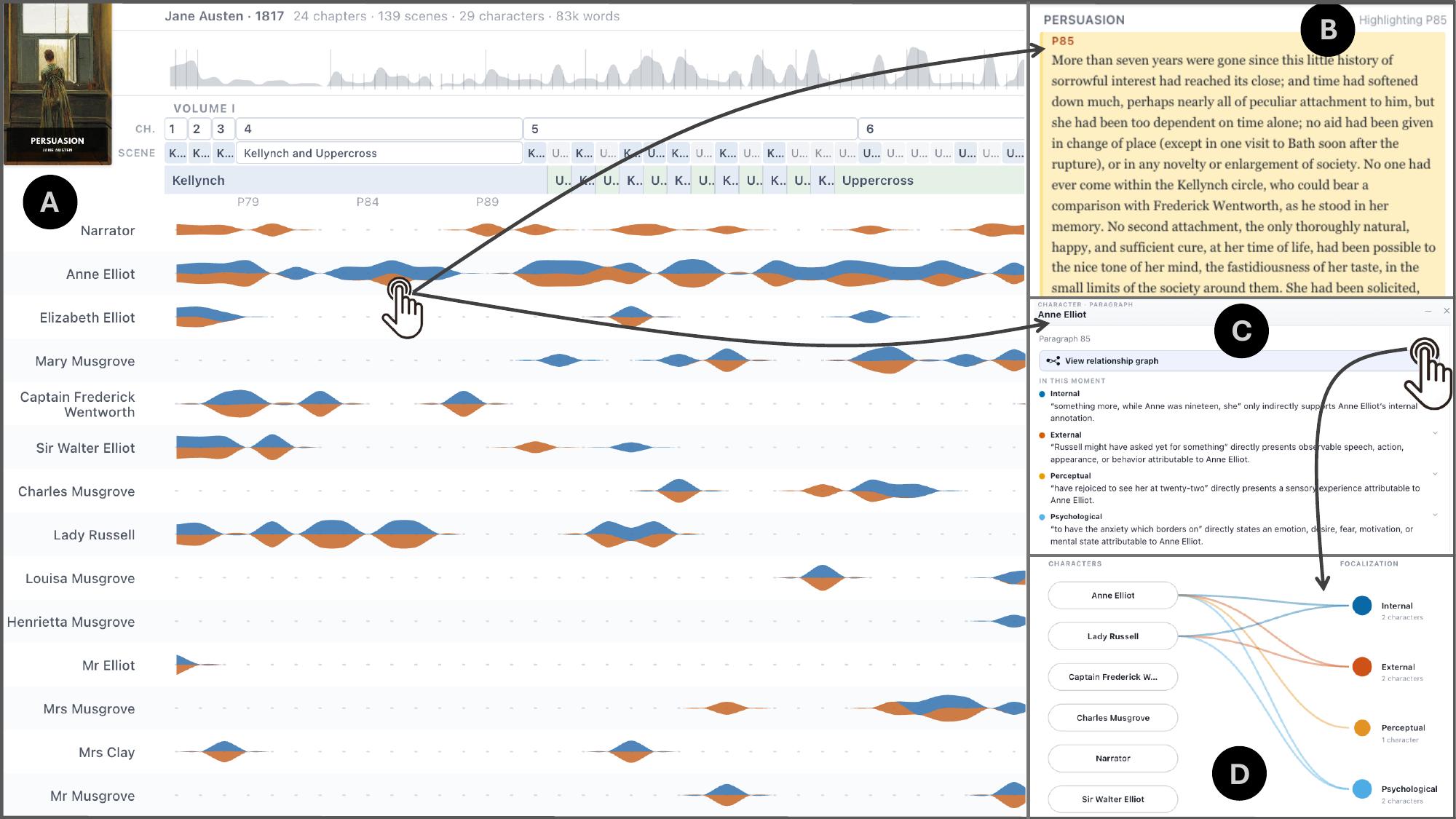}
    \caption{\textbf{Persuasion (1818) by Jane Austen visualized in FocalLens.} (A) The tool uses a streamgraph visualization in which blue represents internal (inner thoughts of a character) and orange external (outward action of a character) focalization. The main visualization is a timeline of the story whose rows and columns represent participating characters and related paragraphs, scenes, or chapters. Selecting specific points in the streamgraph connects users to corresponding passages in a reading interface (B), which highlights the text. (C) The selection also opens a brief explanation of the focalization types and facets for the selected character and text. (D) A widget presents a relationship graph showing the connections between characters and their associated focalization types and facets for the selected passage.}
    \label{fig:focallens-teaser}
\end{figure*}

To test this hypothesis, we explore the potential of visualizing perspective changes in a narrative through an underexplored narrative component named \textbf{focalization}. We start by providing a theoretical foundation of focalization, identifying different types and facets of focalization~\cite{rimmonkenan2002narrativefiction, genette1980narrativediscourse}. This includes \textit{internal} focalization, where a story is narrated through a character's internal thoughts. \textit{External} focalization is the opposite of internal focalization, capturing cases where a story is told from a character's external activities. Focalization facets provide more depth in understanding perspective, including how psychological (cognitive and emotive), perceptual (space and time), and ideological (norms and evaluation) aspects can shape the narration of a story. 
We then conducted formative interviews with four literary scholars to understand their perspectives and challenges regarding focalization.  We showed participants existing visualization techniques for narrative analysis as probes~\cite{iStoryline, hoque2023portrayal, DBLP:journals/tvcg/KimBISGP18}. Participants appreciated the visualizations and mentioned that a similar tool built upon the theory of focalization could offer a common vocabulary for analyzing focalization. They also provided requirements for understanding focalization in the context of a story.

Based on these findings, we designed \textbf{FocalLens} (Figure~\ref{fig:focallens-teaser}), an interactive visual interface for analyzing focalization. The interface is powered by an LLM-based pipeline, developed based on experiments on a set of reference annotations containing focalization labels for 300 paragraphs from four literary works. The primary visualization is a stream-graph-like chart that shows the distribution of focalization types and facets for characters over time. We provide flexible interactions to select types and facets of interest, modify the chart axis, and move between different hierarchies of a novel (e.g., scenes and chapters). A character-focalization bipartite graph further shows relations among characters, focalization types, and facets for the selected text region. Finally, an explanation module enables users to understand the rationale behind LLM-generated labels.

We evaluated \toolname{} with 13 literary scholars in a user study. We asked participants to analyze two stories from a set of 12 well-known fictional novels using two tools: FocalLens and Story Ribbons~\cite{yeh2025story}. In addition, we asked seven participants who were also creative writers to analyze one of their early drafts in \toolname{}. We found that \underline{\toolname{}} facilitated a large number of unique insights about perspectives and narration compared to Story Ribbons (\underline{38} vs 9). Participants also found more insights about the overall writing style while using \underline{\toolname{}} (\underline{9} vs 0). In contrast, while using \underline{Story Ribbons}, participants found more insights about plot structure (12 vs \underline{22}), character interaction (4 vs \underline{12}), and character traits (5 vs \underline{14}), compared to \toolname{}. We conclude by outlining design implications for combining the complementary capabilities of the tools and, more generally, for the future of visualizing narratives. Overall, this work offers the following contributions:
\begin{itemize}
    \item A \textbf{formative study} grounded in theoretical formalization with four literary scholars to derive design requirements for visualizing focalization in literary narratives.
    \item \textbf{\toolname{}}, an interactive visual interface that represents focalization across narrative progression and multiple levels of granularity while linking visual patterns to the underlying textual evidence.
    \item An \textbf{LLM-based annotation pipeline} for extracting paragraph--character focalization annotations, comparing 10 LLMs using a custom-built 300-paragraph dataset.
    \item A \textbf{user study} with 13 scholars (seven of whom were also creative writers) for comparing FocalLens with Story Ribbons~\cite{yeh2025story} to establish the insights and workflows supported by \toolname{} in contrast to the current state-of-the-art. 
\end{itemize}

%% file: sections/02-background.tex
\section{Background: Narrative and Focalization}
\label{sec:background}

A narrative is a representation of events organized over time to convey a story, experience, or sequence of actions~\cite{prince2003dictionarynarratology, bal2009narratology}. Researchers have proposed several ways to study a narrative. Earlier theories conceptualized narrative through event order, characters, time, and location~\cite{prince2003dictionarynarratology}. It was Genette~\cite{genette1980narrativediscourse} who first introduced the concept of perspective in a narrative through constructs like narrator and focalizer. According to Genette, the construct \textit{narrator} asks \textit{``who speaks?''} in a story whereas the construct focalizer asks \textit{``who sees or perceives?''}. For example, consider the following simple text:

\begin{quote}
    ``Samantha was walking in the room.
    She looked at the broken vase and felt a sudden wave of guilt.''
\end{quote}

A third-person narrator is speaking in this event. Samantha is not narrating the events. However, it is Samantha who sees (i.e., looked at the broken vase) and perceives the events (i.e., felt a sudden wave of guilt). Thus, Samantha is the focalizer here. Note that the same character can be both the narrator and focalizer, but they need not be. A third-person narrator, for example, may report a child's fear, a soldier's confusion, or a dying man's final wish without being any of those characters. 

\subsection{Types of Focalization}
Rimmon-Kenan distinguishes focalization by the position of the character relative to the story world~\cite{rimmonkenan2002narrativefiction}. The two types are as follows:

\subsubsection{Internal focalization}
With internal focalization, a reader has access to the direct and inner perceptions, knowledge, and emotions of a character. The narrative is anchored in a character inside the story world. George R.R.\ Martin opens the first chapter of \emph{A Game of Thrones}~\cite{martin1996gameofthrones} from the viewpoint of Bran Stark, a seven-year-old boy. Everything in the chapter is reported as Bran \fixx{perceives and understands it}:
\begin{quote}
``Bran rode among them, nervous with excitement. [\ldots] He had taken off Father's face, Bran thought, and donned the face of Lord Stark of Winterfell.''
\end{quote}
\noindent
Thus, Bran \fixx{is the focalizer and the chapter} is internally focalized, since we have direct access to his perceptions, knowledge, and emotions. \fix{Ned} Stark's expression is interpreted through Bran's viewpoint.

\subsubsection{External focalization}
With external focalization, the perspective is available only through indirect and outward behavior of a character. The text reports outward behavior without entering a character's mind~\cite{rimmonkenan2002narrativefiction}. Stephen Crane introduces the cook in \emph{The Open Boat}~\cite{crane1898openboat_gutenberg} through observable actions:
\begin{quote}
``The cook squatted in the bottom and looked with both eyes at the six inches of gunwale which separated him from the ocean.''
\end{quote}
\noindent
The sentence reports only the cook's posture and gaze. It does not tell us what the cook actually thinks or feels. The reader receives the same information that a detached observer could perceive.

\subsection{Facets of Focalization}
The types above explain \emph{where} the focalizer is positioned (internal or external). They do not explain \emph{what aspect of experience} the passage foregrounds. Rimmon-Kenan addresses this through three \emph{facets} of focalization: \emph{perceptual}, \emph{psychological}, and \emph{ideological}~\cite{rimmonkenan2002narrativefiction}. These facets help distinguish whether a passage mainly limits what can be sensed, reveals what a character knows or feels, or presents the norms and judgments through which events are understood. A passage may show more than one facet at once, but one is often more prominent than the others.

\subsubsection{Perceptual facet (space and time)}
The perceptual facet concerns sensory perception of the focalizer (seeing, hearing, noticing, etc.).   Stephen Crane opens \emph{The Open Boat} by making this facet explicit:
\begin{quote}
``None of them knew the color of the sky. Their eyes glanced
level, and were fastened upon the waves that swept toward them.''
\end{quote}
\noindent

This is the perceptual facet because the passage describes what the focalizer can see, hear, or otherwise perceive, and from where.

\subsubsection{Psychological facet (cognitive and emotive)}
The psychological facet concerns the cognitive state, emotion and feelings of a focalizer. The following excerpt from \emph{Harry Potter and the Philosopher's Stone} by J.K.\ Rowling exemplifies the psychological facet:

\begin{quote}
```Mum?' he whispered. `Dad?' They just looked at him, smiling. [\ldots] Harry was looking at his family, for the first time in his life. [\ldots] He had a powerful kind of ache inside him, half joy, half terrible sadness.''
\end{quote}
\noindent
This passage is shaped by Harry's feeling: joy, grief, and, longing. This is the Psychological facet. It also contains the perceptual facet since readers can perceive what Harry is doing in the scene.

\subsubsection{Ideological facet (norms and evaluation)}
The ideological facet \fix{concerns the norms, values, and judgments of a focalizer}. For example, consider the following excerpt from \emph{Persuasion} by Jane Austen:
\fix{
\begin{quote}
``An admiral speaks his own consequence, and, at the same time, can never make a baronet look small.''
\end{quote}
Here, the narration reflects Sir Walter's concern with social status and his belief that an admiral's rank should not overshadow his own.} Thus, the passage captures the ideological facet through Sir Walter's values and judgments.

%% file: sections/03-related-work.tex
\section{Related Work}
\label{sec:related-work}

\subsection{\textbf{Computational Approaches for Literary Analysis}}
Computational approaches to literary analysis have a long history in the field of Digital Humanities, where researchers use computational methods to examine literature and historical documents. Writers often have distinct literary styles, and computational tools can help scholars identify patterns in language use and writing styles. Voyant Tools~\cite{voyant}, Google Ngram Viewer~\cite{michel2011quantitative}, Hedonometer~\cite{reagan2016emotional}, and Wordle~\cite{DBLP:journals/tvcg/ViegasWF09} are examples of tools that support text analysis through word frequencies, linguistic trends, and sentiment patterns. 

Beyond traditional text analytics, NLP research has introduced computational approaches for extracting and analyzing narrative information. One of most influential works is BookNLP~\cite{bamman2020coreference}, an open-sourced NLP library that can support character identification and coreference resolution in book-length documents. Kim et al.~\cite{kim2020time} constructed a dataset of literary time references and developed models to infer the time of day of events in novels, including passages without explicit temporal cues. Pial et al.~\cite{DBLP:conf/emnlp/PialAPKS23} combined semantic similarity with sequence alignment to match passages from novels with film scripts, supporting the analysis of narrative content retained or omitted during adaptation. Eisenberg and Finlayson~\cite{eisenberg2016automatic} developed classifiers for narrative point of view and diegesis, while Brooke et al.~\cite{brooke2017freeindirect} used lexical style models to analyze free indirect discourse.

Most relevant to our work is Hicke et al.'s experiments in identifying focalization from Stephen King novels~\cite{hicke2025says}. GPT-4o achieved the highest accuracy in detecting focalization categories. However, their work focuses on model performance, while we focus on how focalization can be used to represent narratives visually and how that can enable new analytical capabilities and insights for literary scholars. We also extended their LLM experiments by developing a new ground truth dataset since their dataset is not publicly available and is limited to only one author.

\begin{figure}
    \centering
    \includegraphics[width=\linewidth]{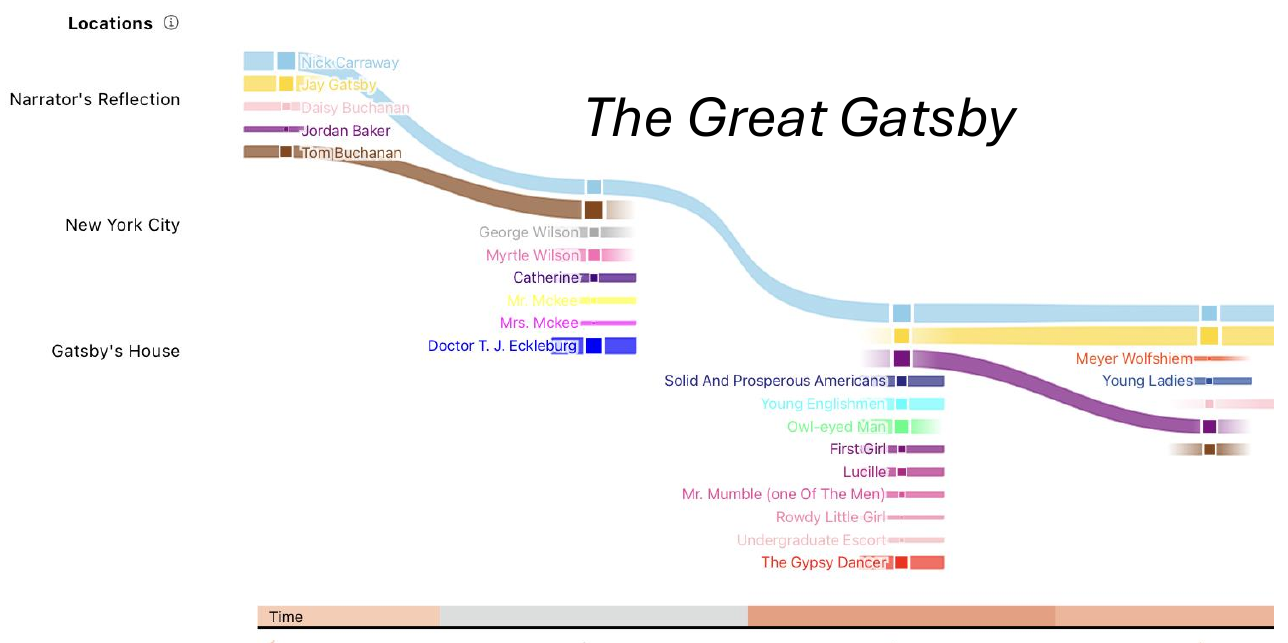}
    \caption{\textbf{A screenshot of the Storyline visualization in the Story Ribbons tool~\cite{yeh2025story}.} The x-axis represents narrative time and the y-axis represents narrative locations. Characters are grouped based on their co-occurrence in different narrative locations. Image credit: \url{https://catherinesyeh.github.io/story-demo/}.}
    \label{fig:storyline}
\end{figure}

\subsection{\textbf{Visualizing Narratives}}
Visualizing the internal mechanics of narratives is a well-established subfield within data visualization and HCI~\cite{janicke2015close}. 
Starting from generic text visualizations such as word clouds~\cite{hearst2019evaluation, DBLP:journals/cga/CuiWLWZQ10}, TextFlow~\cite{DBLP:journals/tvcg/CuiLTSSGQT11}, and ThemeDelta~\cite{DBLP:journals/tvcg/GadJGEEHR15}, researchers have proposed several visualizations for representing character and scene dynamics in a narrative. The Storyline visualization~\cite{DBLP:journals/tvcg/TanahashiM12, DBLP:conf/gd/GronemannJLM16, DBLP:journals/tvcg/LiuWWLL13, dibartolomeo2026survey} is perhaps the most popular and well-regarded visualization in this area (Figure~\ref{fig:storyline}). It maps character co-occurrences and relationship dynamics over a temporal axis. The visualization does not represent the chronological order of events in a story, but rather how the events from the story are actually told in the text or movie (i.e., narrative). The Story Curve~\cite{DBLP:journals/tvcg/KimBISGP18} visualization represents both story and narrative timelines. StoryPrint~\cite{DBLP:conf/iui/WatsonSSGMK19} utilizes circular timelines to concurrently map scene progression and character sentiment. Portrayal~\cite{hoque2023portrayal} visualizes different character indicators (e.g., sentiment, actions, adverbs, etc.) in heatmaps and word clouds. Most recently, Story Ribbons~\cite{yeh2025story} extended the Storyline visualization and leveraged LLMs to automatically extract and visualize character, location, and thematic trajectories from unstructured literary text. These systems demonstrate the value of aggregate narrative views, while
leaving open how an interpretive component can be examined consistently
across local passages, scenes, and broader narrative structure.

However, none of the existing works visualize focalization, an important narrative component in literary theories~\cite{rimmonkenan2002narrativefiction}. We have noticed a lack of research for visualizing complex narrative components (e.g., speech, cause and effect relations, narration, character outlooks) that go beyond showing the presence of characters and locations in a narrative. 
We believe capturing and visualizing complex narrative components can create a new generation of literary tools that can facilitate nuanced insights about narratives. This paper takes a  step toward that goal.

\subsection{\textbf{Writing Support Tools and Visualization}}
Beyond supporting narrative analysis, visualizations can also assist writers in developing and refining their work. Contemporary writing support tools provide assistance through grammar correction, text completion, and summarization~\cite{DBLP:conf/chi/0002GCSRSVWZAAB24}. More recently, the integration of LLMs has introduced new opportunities for interactive writing support, where visual interfaces allow writers to guide and refine AI-generated content~\cite{DBLP:conf/chi/HoqueMGSCKE24}. For example, TaleBrush~\cite{talebrush} allows users to sketch a character's fortune line on a canvas to procedurally generate story text. Visual Story Writing~\cite{DBLP:conf/uist/MassonZC25} introduced a visual interface that allows writers to manipulate story elements, such as characters, locations, and timelines, to guide LLM-assisted story generation.

We primarily situate \toolname{} as a literary analysis tool. It does not have an active writing interface. However, it has implications for writing support tools. Most importantly, literary scholars are often writers too, as depicted in our participant pools. Thus, the analysis they perform in tools like \toolname{} can inform their writing too. Prior research has shown such indirect influence on writing by narrative analysis tools~\cite{DBLP:conf/ACMdis/HoqueGE22, hoque2023portrayal}. We report on how writers used \toolname{} in the user study and discuss the future of writing support tools integrating narrative components as a feedback mechanism. 

%% file: sections/04-design-of-focallens.tex
\section{Formative Study}
To understand how literary scholars view and use focalization, we conducted formative interviews with four scholars (P1-P4). We recruited the participants by reaching out to the English and Creative Writing departments at our university in the U.S. All participants are literary scholars with active research portfolios on literary style in novels and poetry. They had formal higher education training in English or other literary majors, with experience ranging from 5 to 20 years. Two participants (P2, P4) identified themselves as creative writers too. These two professions are overlapping~\cite{hoque2023portrayal}. Three participants were female, and one was male. Participants received \$30 for their time. The interviews were conducted over Zoom.

The interviews were divided into three parts. In the first part, we asked participants to share their experiences with literary analysis. We also asked about their creative writing experience, if any. In the second part, the interview administrator (the first author) shared their screen and presented a slide deck with different focalization types, facets, and examples, similar to \autoref{sec:background}. We then asked participants to share their thoughts about the concepts presented and how they relate to their work. We also asked participants in which scenarios focalization could become an important literary tool and the challenges they might potentially face in those scenarios. Finally, participants brainstormed with the administrator for a potential solution that might visualize focalization. To elicit discussion, we presented screenshots of some popular visualizations representing narratives, including iStoryline~\cite{iStoryline}, Portrayal~\cite{hoque2023portrayal}, and Story Curves~\cite{DBLP:journals/tvcg/KimBISGP18}.

\subsection{Findings}

\subsubsection{Difficult to Operationalize}
Participants found focalization to be a familiar concept but difficult to operationalize. Participants regarded Genette's work and the types and facets we presented as foundational, but mentioned that it can be difficult to define, teach, and apply them systematically. P2 said, \textit{``I have known about focalization, but with different names like lenses, perspectives, point of view, etc. I think I have used [analysis] [\ldots] similar to you internal focalization, without really using the term.''}  P1, who teaches digital humanities and narrative structures, said \textit{``This is a very complex thing to explain and teach in my class. I try to teach the goal and intuition behind it without going too deep into it.''} In light of this discussion, two participants mentioned that the proposed tool could establish a common ground for conceptualizing and applying focalization, which can motivate scholars to think about focalization more actively when analyzing texts.

\subsubsection{Subjective Nature of Focalization}
All four participants expressed that whether a text is external or internal is subjective to the interpretation of the reader. P1 said, \textit{``Well it's a bit tricky to put tags on some texts. They can be interpreted both ways. It is the argument for the tag that is important.''} Participants were unsure how a tool could capture the argument or opinion from a scholar or writer but emphasized that this is an important dimension of focalization.

\subsubsection{Difference between Writing and Analyzing}
Participants who were both scholars and writers (P2, P4) found focalization more relevant to literary analysis than to the writing process. P2 noted, \textit{``When I write, I do not really think about in that (focalization) level.''} However, they suggested some useful applications of focalization for creative writing, especially as a lens for revision and reflection rather than directly influencing writing. Participants also emphasized that critically reading literary works helps writers develop their own styles. P4 explained, \textit{``A big part of writing is just reading. You need to read, appreciate, and analyze others' work to become a mature writer.''} The proposed tool can therefore help writers examine how other authors use shifts in focalization to convey character experiences, emotions, and narrative tension. It can also support reflection on their own drafts by helping them examine whether the intended focalization and access to characters' inner experiences are maintained in the narrative.

\subsubsection{Preference for the Proposed Tool}
The brainstorming discussion generated several recommendations for the tool. All participants responded positively to the visualizations of existing tools, using terms such as ``interesting'', ``amazing,'' and ``cool''. They compared the visualizations to distant reading tools~\cite{moretti2013distant}, which provide broad overviews of narratives and help identify areas of interest. Participants appreciated the character-centric design of these visualizations, noting that it made them easy to read. Among the presented approaches, participants preferred the Storyline visualization because it clearly represented narrative structure without excessive complexity. P2 remarked, \textit{``this [Storyline] is interesting. Maybe your visualization can be something like that.''}

Two other requirements were recurrent: 1) making the visualizations easy to read and 2) providing evidence for the visualizations. While participants were enthusiastic about the visualizations, they mentioned that they can be overwhelming at times and provide a lot of data to comprehend (P1-3). For the second requirement, participants mentioned that they would not trust the visualizations without seeing the actual texts and interpreting them in the context of the texts (P1-4). 

\section{FocalLens: Visualizing Focalization}
\label{sec:design}

Based on the review of the literature and formative interviews, we developed four design goals that drive the development of the tool. They are as follows:

\begin{itemize}
    \item[DG1] \textbf{A Technology Probe for a Shared Vocabulary of Focalization.} A key finding from the formative interviews was that scholars understand focalization intuitively but lack resources to apply it systematically in practice. The theories~\cite{genette1980narrativediscourse, rimmonkenan2002narrativefiction}, combined with appropriate data visualization, can establish a shared standard for scholars and writers to conceptualize focalization. Thus, we decided to adopt the focalization types (internal and external) and facets (perceptual, psychological, and ideological) as the primary data attributes that we want to visualize in the proposed tool~\cite{rimmonkenan2002narrativefiction}. Since, to our knowledge, there are no existing tools for analyzing focalization, we aim to develop the tool as a \textit{technology probe}~\cite{10.1145/642611.642616}, a lightweight and acceptable system intended to investigate user needs, evaluate technological possibilities, and inform future designs. We adopted this approach to explore the potential benefits and limitations of visualizing established focalization categories and explore how the process can be made more customizable for users.
    
    \item[DG2] \textbf{Adopting Storyline Visualization.}  Participants found the Storyline visualization~\cite{dibartolomeo2026survey} to be most interesting and intuitive. Thus, we decided to extend the Storyline visualization and encode focalization into it.  Additionally, participants mentioned that the visualization can be overwhelming in some cases, especially with the amount of information it provides. Participants suggested keeping the proposed visualization easy to read and understand. We also noticed similar requirements in previous research~\cite{hoque2023portrayal, yeh2025story, iStoryline}. Literary scholars and writers are not visualization experts, so the learning curve for the tool should be low. They are mostly interested in seeing the narrative patterns, not learning a complex visualization. 
    
    \item[DG3] \textbf{Interactive Exploration of Narrative Hierarchy.} Like any visual analytics system, interactions could help users manage the complexity of the visualization and gain new insights. Interactions in visual narrative analysis tools are typically useful for free-form exploration, building narrative hypotheses, reducing visual clutter, focusing on specific narrative points, and identifying low-level patterns that are not available in an overview panel~\cite{hoque2023portrayal, DBLP:journals/tvcg/KimBISGP18, yeh2025story}. This translates to support for moving between paragraph-, scene-, and chapter-level visual representations, selecting specific stories and characters, and adding/removing different visual encodings. Thus, our tool should support these types of interactions. 
    
    \item[DG4] \textbf{Support Evidence-based Workflow.}
    A critical feedback we received from the formative interviews was that the rationale for assigning a focalization category is very important to participants, likely even more than the category itself. They want to check the correctness of the representations by cross-referencing with the text. Most often, this approach stems from their academic backgrounds (degrees in English or Creative Writing), which train them in ``close reading''~\cite{richards1929practical, moretti2013distant}, a technique for critically reading and analyzing individual excerpts to interpret the underlying meaning.
    Previous research also suggests that literary scholars and writers approach narrative analysis tools with an evidence-seeking approach~\cite{hoque2023portrayal, DBLP:conf/ACMdis/HoqueGE22, yeh2025story}.  In addition, we aim to use LLMs to extract the focalization types and facets from the content of a text. Previous research suggests that scholars and writers are concerned about the limitations of LLMs in these tools and, more broadly, for creative purposes~\cite{li2024disclosure, chakrabarty2024creativity}. Story Ribbons~\cite{yeh2025story} provided explanations generated by the LLMs along with the labels for improving trust and transparency. Thus, our tool should connect the visualization with the text and provide appropriate rationale for any focalization labels assigned to a text.
\end{itemize}

\begin{figure*}[!ht]
    \centering
    \includegraphics[width=0.92\linewidth]{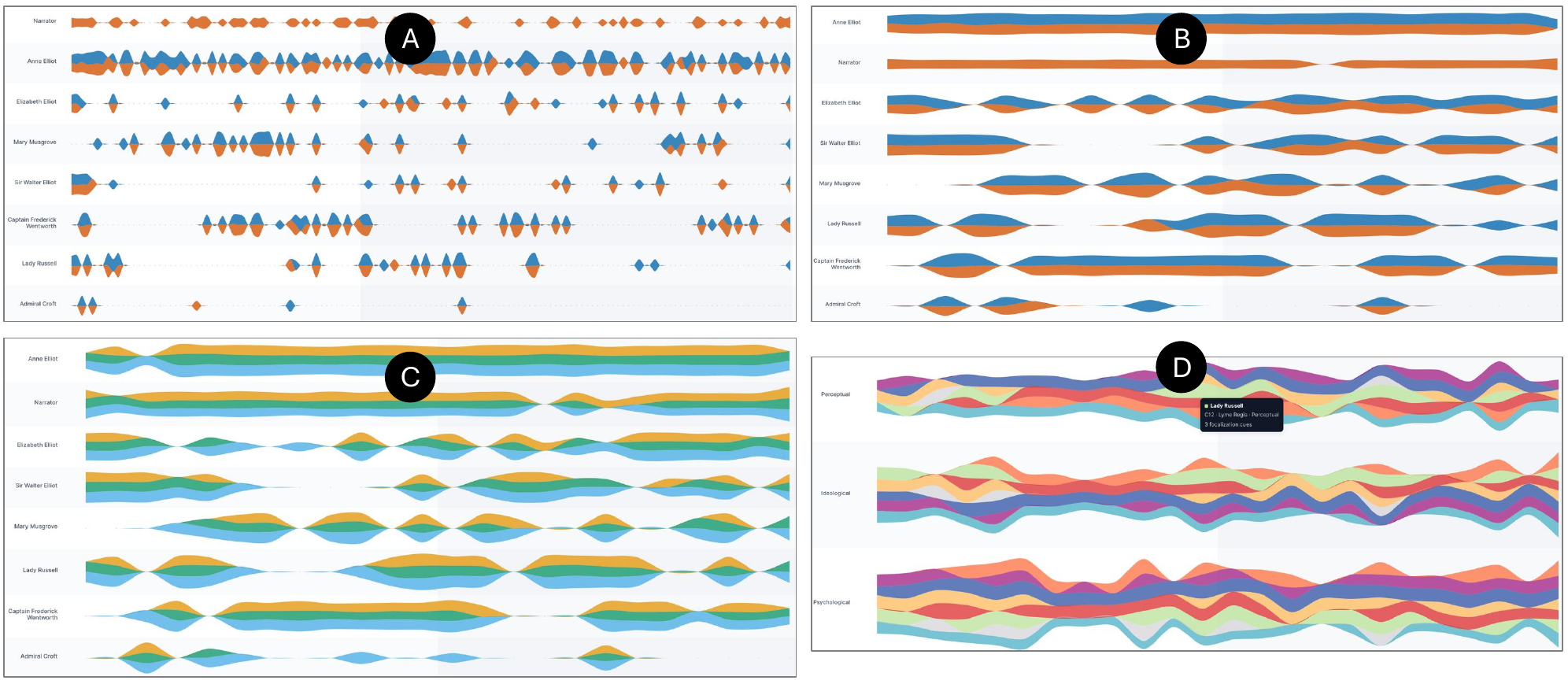}
    \caption{\textbf{Different views in \toolname{}. This example uses \textit{Persuasion} (1818) by Jane Austen.} A) This streamgraph represents focalization types for the characters in the scenes. Blue represents \textit{internal} and orange represents \textit{external}. \fix{The width of each stream represents the number of paragraphs in a scene in which the character received an internal or external focalization label.} B) Similar view but x-axis now represents \textit{chapters}, instead of \textit{scenes}. C) Same streamgraph now visualizing focalization facets (perceptual: yellow, ideological: green, and psychological: light blue). D) Finally, the y-axis can be altered to place focalization types and facets. In this example, three facets are placed in three rows. Each colored stream now represents a character in the story.}
    \label{fig:streamgraph-types}
\end{figure*}

\begin{figure*}[!ht]
    \centering
    \includegraphics[width=0.85\linewidth]{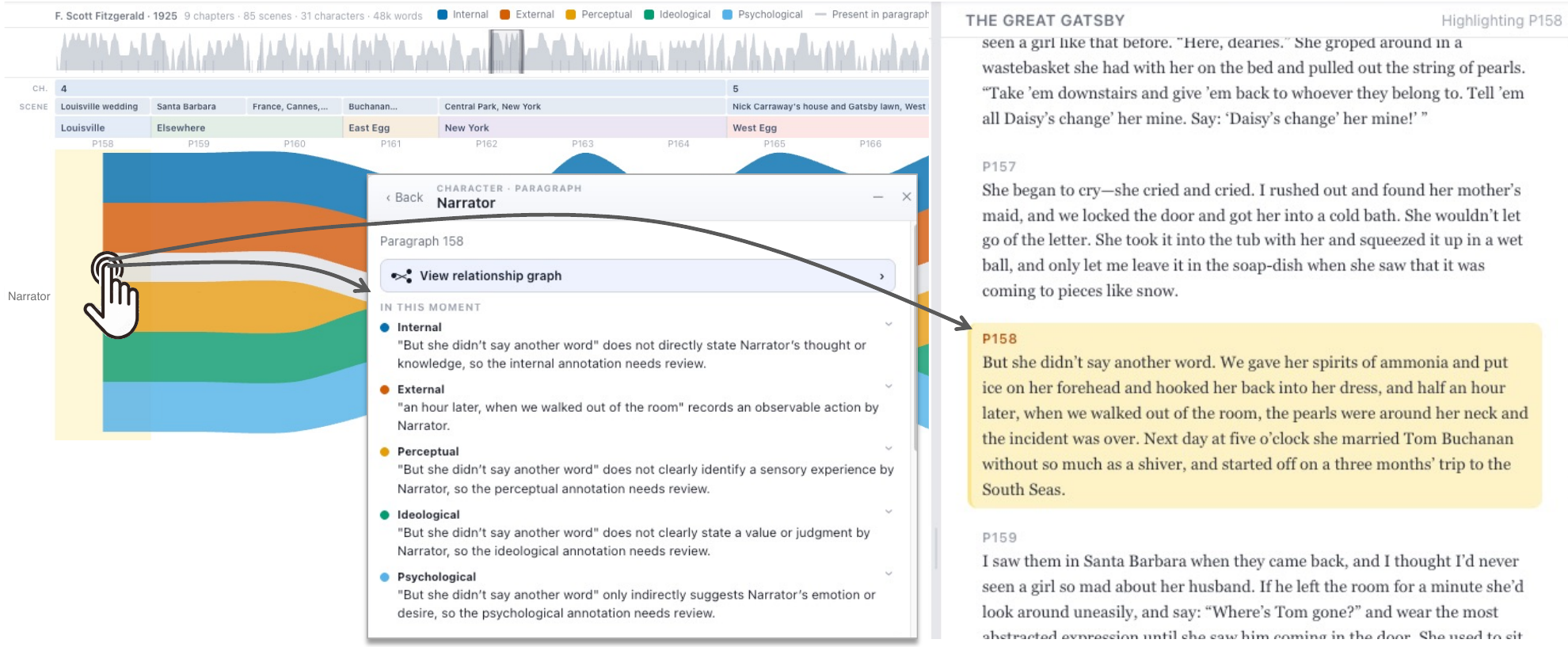}
    \caption{\textbf{Visualization, explanations, and original source.} On clicking a particular point in the streamgraph, an explanation module and a reading interface with the original source text is displayed. The explanation module provides rationale for the assigned focalization labels.}
    \label{fig:interaction}
\end{figure*}

Guided by the design goals introduced above, we developed FocalLens, an interactive web-based system for exploring focalization patterns in literary narratives (DG1). The interface consists of three primary components: the Focalization Visualization, the Source Text Panel, and the Interaction Control Bar, complemented by a contextual explanation module (\autoref{fig:focallens-teaser}). In the following, we describe the visual interface of FocalLens.

\subsection{Focalization Overview and Visual Encoding}
\label{sec:interface}

Per DG2, we wanted to integrate focalization related variables into the storyline visualization. However, limited visual channel was a challenge. The storyline visualization places time on the x-axis and locations on the y-axis. Each colored ribbon represents a character, moving through time and locations. Adding one more channel for focalization will make the visualization cluttered and complex. We made the following design decisions to address this challenge:

\begin{itemize}
    \item Mapping time on the x-axis to maintain the intuitive nature of time flowing from left to right, as depicted in almost all prior solutions for visualizing narratives.
    \item Mapping locations to an annotation layer instead of y-axis to free visual channels for characters and focalization. This is similar to how Portrayal encodes locations~\cite{hoque2023portrayal}.
\end{itemize}

Based on these design decisions, we adopted a streamgraph-style visualization that mimics the flow-like design of storyline. Figure~\ref{fig:streamgraph-types} presents different ways to modify the streamgraph. The default view (Figure~\ref{fig:streamgraph-types}A) shows the characters on the y-axis and narrative unit of time (e.g., paragraphs, scenes, or chapters) on the x-axis. Colored segments indicate the focalization types (internal and external) associated with each character. Users can easily change the narrative unit. For example,  Figure~\ref{fig:streamgraph-types}A and B are the same visualizations, but the former uses scenes and the latter uses chapters on the x-axis. This helps users manage the level of detail in the visualization (DG3). Users can flexibly change the focalization types and facets they want to see in the streamgraph. For example, Figure~\ref{fig:streamgraph-types}C shows the variation in three facets of focalization (perceptual, ideological, and psychological), changed from the types in the previous two streamgraphs. \fix{The colors representing focalization types and facets remain consistent across all views.} Users can also select characters of interest to reduce visual clutter.
Finally, users can also change the y-axis from a character-oriented view to a focalization-oriented view. For example, Figure~\ref{fig:streamgraph-types}D is an alternative view of Figure~\ref{fig:streamgraph-types}C. \fix{In the latter view, the three facets are arranged along the vertical axis, while colors distinguish individual characters.} This alternative representation allows users to compare how characters are associated with a particular focalization category. In all views, annotations are placed on top of the visualizations (Figure~\ref{fig:interaction}).

\subsection{Linked Text and Explanations}
\label{subsec:text}

To support evidence-based analytical workflow (DG4), FocalLens coordinates its visual representations with a Source Text Panel and provides a contextual explanation for the assigned focalization categories. Selecting a visual element corresponding to a character and narrative unit highlights the associated paragraph in the Source Text Panel (Figure~\ref{fig:interaction}). 

In addition to highlighting the corresponding passage, selecting a focalization annotation opens an explanation module describing the assigned focalization labels (Figure~\ref{fig:interaction}). The module presents explanations generated by the LLM for the assigned labels. It also includes
contextual information extracted from SparkNotes~\cite{SparkNotes_GreatGatsby}, a popular educational platform that provides study guides, chapter summaries, character lists, and thematic analyses for literature.

\begin{figure}[!ht]
    \centering
    \includegraphics[width=0.8\linewidth]{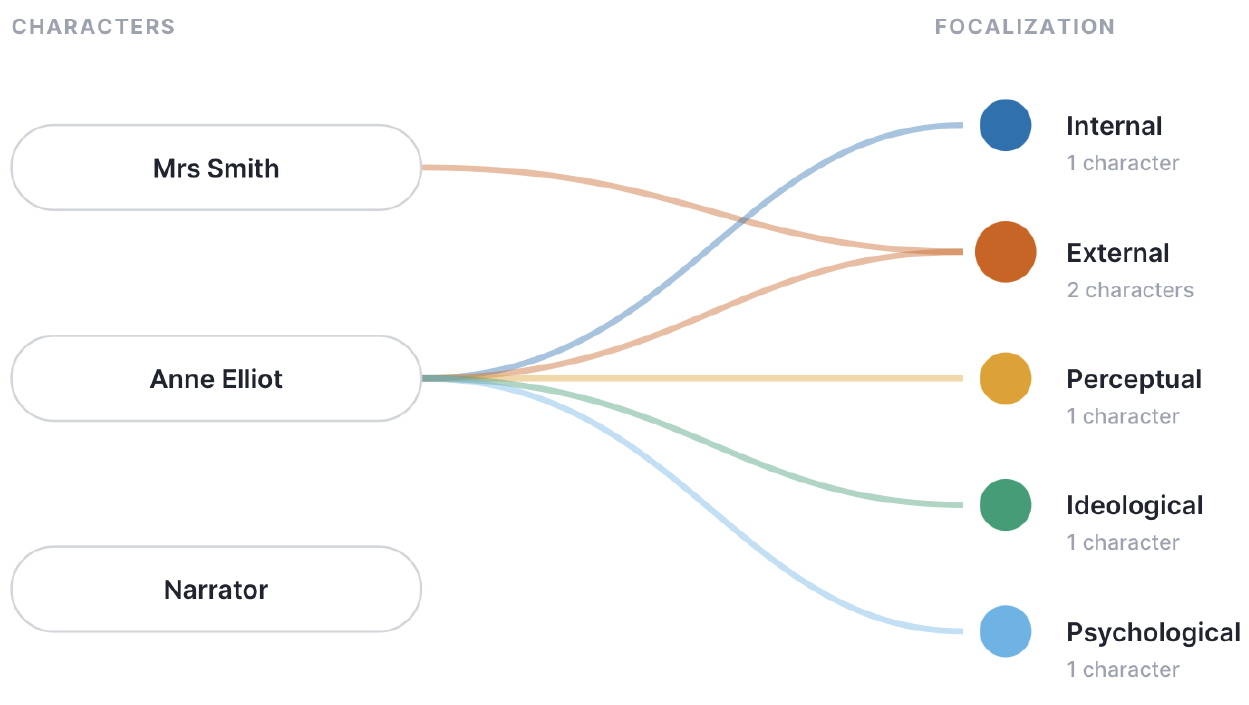}
    \caption{\textbf{The character--focalization association graph.} For a selected narrative unit (paragraph, scene, or chapter), characters appearing in the unit are \fix{displayed} on the left, while focalization types and facets are \fix{displayed} on the right. An edge connects a character to a focalization category when that category is assigned to the character within the selected paragraph.}
    \label{fig:relationship-graph}
\end{figure}

\subsection{Character--Focalization Relationship Graph}
\label{subsec:relationship}

Although the main streamgraph visualization provides an overview of focalization across narrative progression, understanding the associations between characters and focalization dimensions within a specific paragraph may require a more detailed inspection. To support this level of exploration (DG3), FocalLens provides a paragraph-level relationship graph that connects characters to their associated focalization categories (Figure~\ref{fig:relationship-graph}). The graph uses a bipartite representation, with characters on one side and focalization categories on the other.  

\subsection{Implementation Details}
\label{subsec:system-implementation}

FocalLens is an interactive web-based system developed using React and TypeScript, with Vite as the build tool. The visualization components are implemented using D3.js and rendered as SVG elements. \toolname{} is publicly available here: \url{https://focal-lens.github.io/}.

%% file: sections/05-implementation.tex
\section{Narrative Processing and Implementation}
\label{sec:implementation}

FocalLens uses an LLM-powered preprocessing pipeline to convert narrative text into structured focalization data. Following prior LLM-enhanced narrative visualization systems~\cite{yeh2025story}, preprocessing is performed separately from user interaction. The generated annotations, explanations, and narrative structure are stored in JSON format and loaded by the interface for visualization.

\subsection{Model Selection and Evaluation}
\label{sec:llm-eval}

Because focalization involves contextual and subjective interpretation, selecting a suitable language model is important for generating meaningful annotations. We therefore evaluated candidate LLMs before processing the narratives used in FocalLens. 

\paragraph{Dataset construction}
We constructed a set of reference annotations for 300 paragraphs sampled from four literary works: \textit{Ulysses}, \textit{Persuasion}, \textit{The Metamorphosis}, and \textit{The Yellow Wallpaper}. Since multiple characters may appear in a paragraph, the dataset contains approximately 1,000 paragraph--character pairs. Each pair was annotated for five focalization categories: \textit{internal}, \textit{external}, \textit{perceptual}, \textit{ideological}, and \textit{psychological}, along with a binary \textit{focalization} label indicating whether the character is focalized at all (as focalizer or focalized object), equivalent to the logical OR of the five categories; this label is included as the sixth label in all reported metrics.

Two authors developed the annotations in three stages; in each stage, annotating a paragraph means labeling all of its paragraph--character pairs. (1)~\emph{Calibration}: both annotators independently labeled the same 50 paragraphs and resolved disagreements through discussion, refining the guidelines. (2)~\emph{Agreement}: they then independently labeled 100 new paragraphs, reaching a mean Jaccard similarity of 0.94 between their label sets. (3)~\emph{Completion}: the first author labeled the remaining 150 paragraphs, consulting the second annotator and other team members on difficult cases.

\paragraph{Model evaluation}
\fix{Our evaluation included 10 LLMs (GLM4 9B, Mistral Small 3.1, Gemma 3 27B, DeepSeek-R1 32B, Llama 3.1 70B, Qwen3 30B, Qwen3 32B, GPT-4.1, GPT-5, and GPT-5.6) under zero-shot and few-shot prompting settings. We further evaluated multi-agent aggregation and debate configurations. The configurations used two actors and a third adjudicator, all with few-shot prompting. The actors independently annotated each paragraph--character pair, retaining agreed-upon labels. For disagreements, independent aggregation used the adjudicator's independent predictions for majority voting, whereas debate allowed actors to exchange predictions, evidence, and explanations for one round of revision before unresolved disagreements were adjudicated. In both configurations, the final focalization label was derived from the five type and facet labels. Human-annotated reference labels were used only for evaluation and were not shared with the models. The corresponding ablation study is provided in the supplementary material.} We use macro F1 as the primary metric because label prevalence differs across categories, and additionally report Matthews correlation coefficient (MCC) and cell-level accuracy (over all pair--category cells).

\begin{figure}[t]
    \centering
    \includegraphics[width=\linewidth]{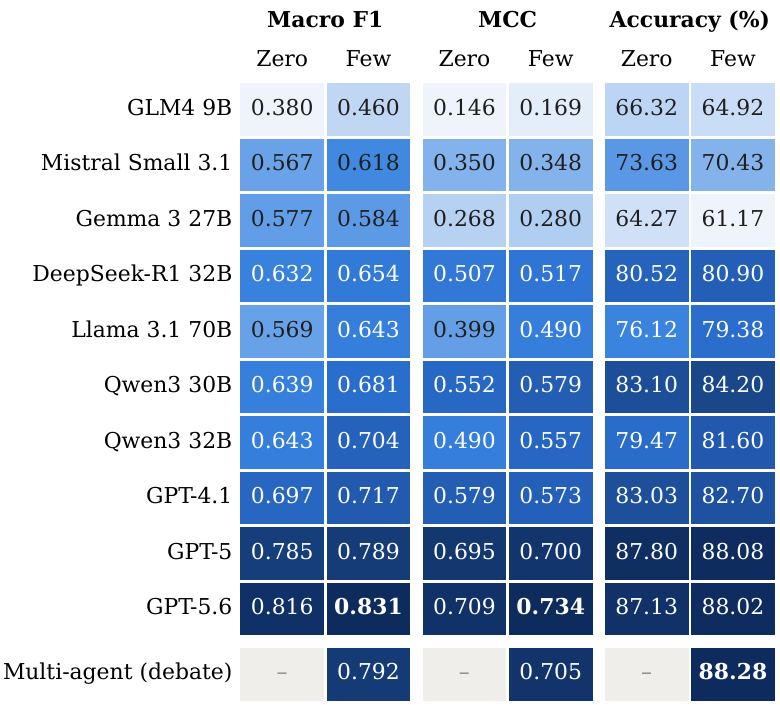}
    \caption{\textbf{Focalization annotation performance.} Zero and Few denote zero-shot and few-shot prompting, respectively. Accuracy refers to cell accuracy. Bold values indicate the best score, and darker cells represent higher scores within each metric. Dashes indicate unreported results. In the multi-agent debate, GPT-5 and GPT-4.1 act as debaters and GPT-5.6 as the adjudicator. The debate setup scores lower than GPT-5.6 alone on macro F1 and MCC. This happens because when the two weaker debaters agree, their answer is accepted without going to the adjudicator. Additional results are provided in the supplementary material.}
\label{tab:annotation-summary}
\end{figure}

As shown in \autoref{tab:annotation-summary}, GPT-5.6 with few-shot prompting achieved the highest macro F1 (0.831) and MCC (0.734). Although the best multi-agent configuration achieved slightly higher accuracy (88.28\%), its macro F1 was lower (0.792). We therefore selected GPT-5.6 with few-shot prompting for narrative preprocessing. These scores measure agreement with our reference annotations rather than absolute literary correctness.

\subsection{Narrative Processing}
\label{subsec:focalization-pipeline}

We organize each narrative hierarchically into chapters, scenes, and paragraphs. For multi-chapter narratives, each chapter is processed independently. Within each chapter, we use an LLM to identify scene boundaries, following Yeh et al.~\cite{yeh2025story}. The model is prompted to detect transitions between scenes, primarily based on changes in narrative location and context. We then retain the paragraphs in each scene and use an LLM to identify unique characters present or referenced in each paragraph. For first-person narratives, the narrator is also represented as a character row to capture their focalization patterns, even when they are not explicitly named in the text.

The basic annotation unit is a paragraph--character pair. For each pair, GPT-5.6 independently predicts \textit{internal}, \textit{external}, \textit{perceptual}, \textit{ideological}, and \textit{psychological} focalization and generates a short explanation grounded in the passage. The categories are non-exclusive and may co-occur.

In addition to predicting focalization categories, the model generates a short explanation grounded in the corresponding paragraph. The displayed explanations are enriched by incorporating analyses and summaries from SparkNotes~\cite{SparkNotes_GreatGatsby}.

%% file: sections/06-evaluation.tex
\section{Evaluation}
\label{sec:eval}
We performed a user study with 13 literary scholars. We decided to primarily test the insight-generation capabilities of \toolname{}. Insight is an important evaluation criterion for visualization and is often seen as the primary outcome of visualization systems~\cite{1626178, 1432690}. This is particularly relevant to this work as scholars are primarily interested in gaining new insights from narrative analysis tools, rather than being more efficient in the task~\cite{hoque2023portrayal, yeh2025story}. Following Saraiya et al.~\cite{1432690}, we define an insight in our context as a distinct observation or discovery that reveals new information or understanding about a narrative, supported by evidence from the visualization and text. The primary research question that we ask is as follows:

\textbf{RQ1:} What new insights does FocalLens facilitate compared to existing solutions for visualizing narratives?

We additionally seek to answer two questions relevant to the usability of \toolname{} and the cognitive load associated with using the tool. They are as follows:

\textbf{RQ2:} How does the perceived usability of FocalLens compare to existing solutions for visualizing narratives?

\textbf{RQ3:} How does the perceived cognitive load of using FocalLens compare to existing solutions for visualizing narratives?

\subsection{Study Conditions}

\begin{enumerate}[label=C\arabic*.]
    \item \textbf{FocalLens:} This is the tool proposed in this paper.

    \item \textbf{Story Ribbons:} We used the  Story Ribbons~\cite{yeh2025story} implementation as the representative of existing solutions for visualizing narratives. The reasons include its: 1) use of storyline visualization, the most well-known solution for visualizing narratives; 2) comparable interactive functions; and 3) use of similar backend technology (LLMs).
    
\end{enumerate}

\subsection{Participants}
We recruited thirteen participants (P1--P13) through personalized email invitations. None of the participants were involved in the formative study. We identified potential participants from publicly available university directories and program webpages based on the relevance of their writing, literary, or teaching background, and contacted them individually. Additional participants were recruited through professional and institutional networks. 

All participants were scholars, but seven identified as writers too. The sample also included an editor (P4) and a creative writing professor (P6). Their areas of expertise spanned fiction, creative nonfiction, poetry, literary criticism, narrative theory, digital humanities, and other forms of narrative analysis and writing. Participants varied in self-reported gender (female = 7, male = 4, prefer not to say = 2) and age (18--24 = 2, 25--34 = 9, 35--44 = 2). Participants received \$30 for their time.

\subsection{Tasks}
Evaluation focusing on measuring insights typically involves open-ended exploratory tasks~\cite{1626178, 1432690}.
Additionally, narrative analysis does not follow a single predefined workflow~\cite{hoque2023portrayal}. Thus, \fix{we adopted an open-ended exploratory task to examine participants' analytical processes and the insights emerging from their interactions. We asked participants to explore one or two pre-loaded stories in the tools within a fixed time range (20 mins) and identify interesting patterns and insights from the tools.} We encouraged them to inspect characters, scenes, focalization patterns, relationships, and other features they found relevant, and to verbalize the observations and questions that emerged.  We additionally collected the System Usability Scale (SUS) after each visualization condition to measure perceived usability and the NASA Task Load Index (NASA-TLX) to measure perceived workload.

\input{table/table2}

\subsection{Procedure}

Each session lasted approximately 60 minutes and was conducted remotely via videoconferencing software (e.g., Zoom). We counterbalanced the presentation order of the two visualization conditions across participants to mitigate potential order effects. Before each condition, we provided a brief tutorial introducing the concepts and visual encodings needed to use that visualization. For FocalLens, the tutorial introduced focalization, its internal and external types, and the perceptual, ideological, and psychological facets represented in the system. We then demonstrated the main visual encodings and interactions using a sample story. For StoryRibbons, we similarly introduced its visual encodings, narrative dimensions, and interactions using a sample narrative. Participants were encouraged to ask questions and try different features of each interface before beginning the corresponding exploration. 

After the tutorial, participants explored the first assigned visualization condition while following a think-aloud protocol. They then completed the corresponding surveys (NASA-TLX and SUS). Participants subsequently explored the second visualization condition using the same general procedure and completed the second survey.

\fix{Since the formative study revealed some applications of the tool for creative writers and our participant pool included seven participants who were both scholars and writers, we decided to explore this dimension of the tool. 
We asked the seven writers to provide samples of their own writing before the study session. We preprocessed those samples and made them available to writers after they completed all tasks and surveys with both conditions. We asked them to reflect on whether the visualization aligned with their intentions, revealed patterns they had not previously noticed, and could inform revisions to their work. This component was designed as a separate case study, conducted after the controlled study, to explore the tool's potential for supporting creative writing.}

Each session concluded with a semi-structured interview focusing on the interpretability and usability of the visualizations. The interviews explored the relationships between visual patterns and the source text, differences between the two representations, and their relevance to participants' writing, literary analysis, or teaching practices. Participants also provided suggestions for future improvements.

\subsection{Analysis}
We categorized the insights reported by participants using a grounded theory approach. We excluded insights generated from participants' own writing drafts, as that part was only performed in \toolname{} and involves a separate task. Two authors first coded the insights in a subset of four session transcripts. The coders then met and discussed different cases and disagreements to develop an insight codebook, outlining different insight types, their definitions, and examples. The rest of the transcripts were analyzed by the first author, in discussion with the second coder and other team members. During this phase, two new insight types were discovered, and some definitions were modified to accommodate emerging cases. The second coder also validated the labels before they were finalized. The semi-structured interview transcripts were analyzed using reflexive thematic analysis. Two authors collaboratively coded the transcripts to thematically group participant feedback.

For the quantitative evaluation of usability (RQ2) and cognitive load (RQ3), we compared SUS and NASA-TLX scores between the conditions. \fix{We report condition means, paired mean differences, 95\% bootstrap confidence intervals, and paired standardized effect sizes. We additionally conducted two-sided Wilcoxon signed-rank tests as nonparametric robustness checks for the SUS and NASA-TLX scores. }

\subsection{Results}
\label{subsubsec:user-study-results}

We present our findings according to the three research questions (RQs):

\subsubsection{RQ1: Differences in Insights}

We found that \toolname{} and the Story Ribbons  support different aspects of narrative analysis. We identified seven types of insights from our analysis (Table~\ref{tab:types}). Insights about story perspectives, focalization, and narration were more frequent in \toolname{}, compared to Story Ribbons (38 vs 9). Participants also identified more insights about writing style (9 vs 0) and scene dynamics (7 vs 3) while using \toolname{}. In contrast, while using Story Ribbons, participants found more insights about the overall plot structure (12 vs 22), character interaction (4 vs 12), and character traits (5 vs 14). Participants mentioned that it was easier to see character groupings and how they move through different scenes in Story Ribbons, which helped them identify insights about character interactions and plot structures (P1, P3-P4, P6, P8, P10-P13). For example, P12 referred to Story Ribbons as ``\textit{more like a map of the story},'' where they could see ``\textit{where they are together}'' and when characters ``\textit{split up}''. In contrast, \toolname{} shifted their focus to story perspectives, focalization, and narration, while still facilitating insights about plot structures and characters (P2-P7, P9-P13). For example, P13 identified a ``\textit{ladder structure}'' in \toolname{} where focalization ``\textit{gets handed off}'' among supporting characters, comparing this progression to ``\textit{a relay race or hot potato}.'' \fix{To further demonstrate how \toolname{} supports literary analysis, we present two detailed case studies drawn from our user study. }

\begin{figure*}[!ht]
    \centering
    \includegraphics[width=0.85\linewidth]{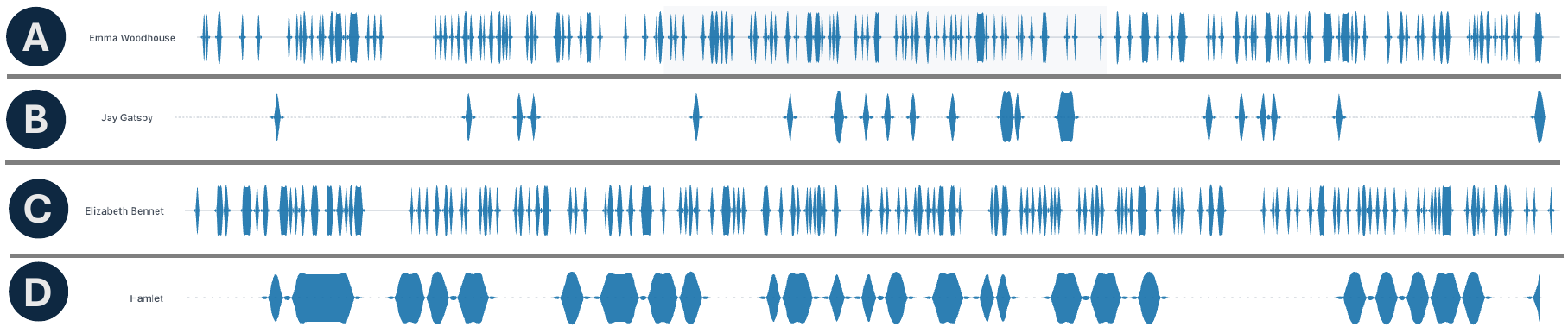}
    \caption{\textbf{Comparing internal focalization across literary works.} The views show the distribution of internal focalization for (A) Emma Woodhouse in \textit{Emma}, (B) Jay Gatsby in \textit{The Great Gatsby}, (C) Elizabeth Bennet in \textit{Pride and Prejudice}, and (D) Hamlet in \textit{Hamlet}. P11 compared these patterns to explore potential differences in narrative styles across authors.}
    \label{fig:comparing-internal-focalization}
\end{figure*}

P11 used FocalLens to compare focalization patterns across literary works (\autoref{fig:comparing-internal-focalization}). P11 began with \textit{Emma}, motivated by prior knowledge of Jane Austen's use of free indirect discourse. P11 selected internal focalization and examined its distribution in \textit{Emma}, then switched to \textit{The Great Gatsby} \fix{to compare focalization patterns across the two novels.} After examining both visualizations, P11 observed: \textit{``Emma has, like, a much more dense focalization than Gatsby.''}

P11 then opened \textit{Pride and Prejudice}, another Austen novel, to examine whether a similar pattern could be observed. The participant further extended the comparison to William Shakespeare's \textit{Hamlet}, examining the internal focalization patterns of the character Hamlet. P11 observed noticeable differences in the distribution of internal focalization across these works, raising questions about how focalization patterns might reflect differences in the narrative styles of these prominent authors.

P11 subsequently connected these observations to a broader literary question: whether patterns of internal focalization could provide additional evidence for characterizing an author's narrative style. P11 concluded that she will consider answering this question through further research, possibly using \toolname{}.

\begin{figure}[!ht]
    \centering
    \includegraphics[width=0.95\linewidth]{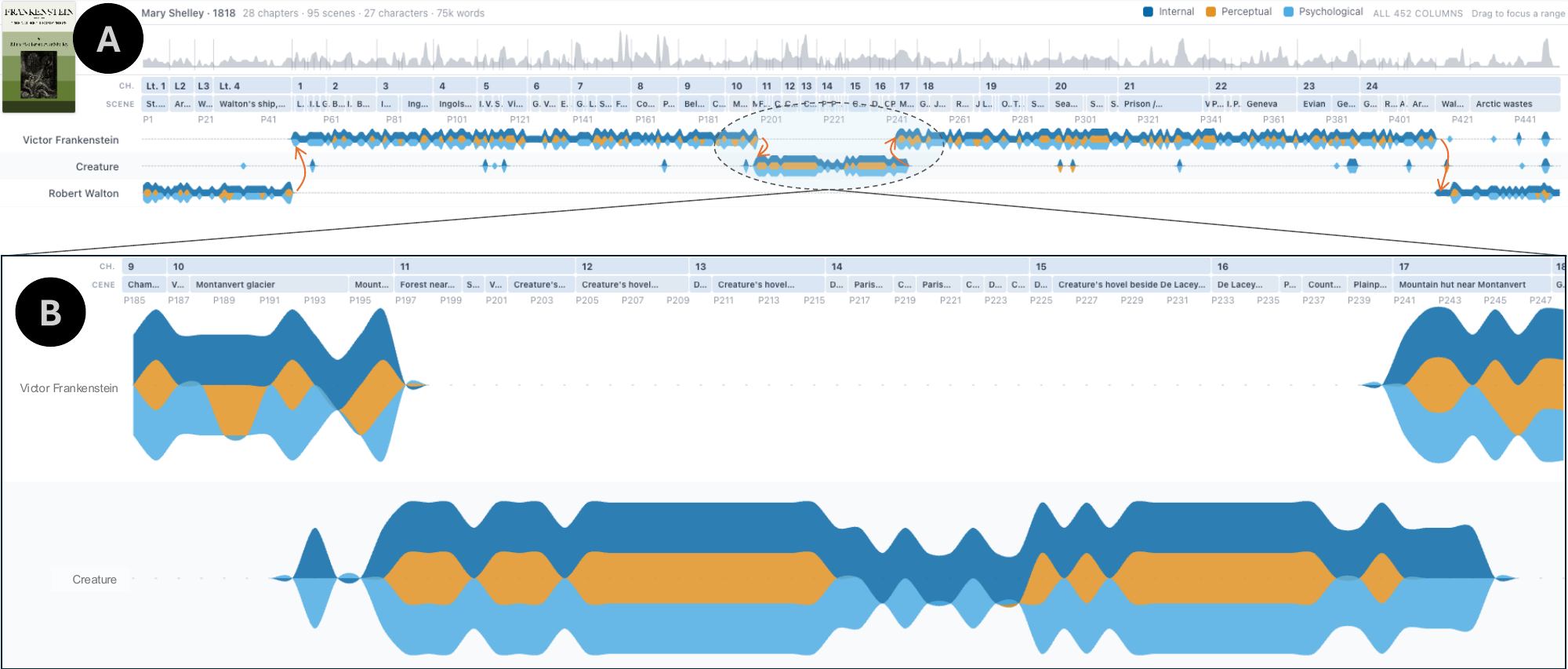}
    \caption{\fix{\textbf{Focalization shifts across nested narratives in \textit{Frankenstein}.} Dark blue, yellow, and light blue represent internal, perceptual, and psychological focalization, respectively. (A) Overview showing Walton (third row) as the primary focalizer at the beginning and end, Victor (first row) through much of the middle, and the Creature (second row) during his inset narrative. (B) Enlarged view showing the shift from Victor to the Creature and back. The relationships between these nested narratives remain implicit.}}

    \label{fig:frankenstein-nested}
\end{figure}

P4 explored \textit{Frankenstein} (\autoref{fig:frankenstein-nested}), observing how Walton serves as a focalizer at the beginning and end, while the Creature becomes the primary focalizer during his inset narrative. P4 appreciated how FocalLens captured these shifts but also noted the complexity of representing nested narrative perspectives:

\begin{quote}
``I like the way that this registers the shifts in [\ldots{}] narrative presence. [\ldots{}] [O]f these three [\ldots{}] focalizers, they do still, like, nest inside each other.''
\end{quote}

\fix{P4 further observed that Victor remains present as the listener and later recounts the Creature's story to Walton, even when his focalization is not explicitly indicated. Similarly, P4 questioned whether Walton's focalization is absent or simply ``not [\ldots{}] the most important one.'' These observations illustrate how FocalLens captures shifts in focalization.}

\subsubsection{RQ2: Perceived usability}
\begin{figure}[t]
    \centering
    \includegraphics[width=\linewidth]{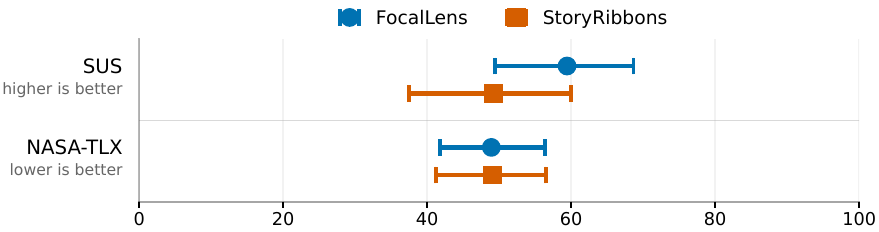}
    \caption{System Usability Scale (SUS) and NASA Task Load Index (NASA-TLX) workload scores for \toolname{} and Story Ribbons. Error bars show 95\% bootstrap confidence intervals.}
    \label{fig:sus-and-nasa}
\end{figure}

\autoref{fig:sus-and-nasa} summarizes the SUS scores. Participants provided higher System Usability Scale (SUS) scores to \toolname{} than  Story Ribbons, but the difference was not statistically significant. \toolname{} received a mean SUS score of $M=59.42$ ($SD=18.68$, 95\% CI $[49.62,\allowbreak 69.42]$), compared with $M=49.23$ ($SD=20.83$, 95\% CI $[37.69,\allowbreak 60.00]$) for Story Ribbons. However, this difference was not statistically significant (Wilcoxon signed-rank test, $W=15.5$, $p=0.064$). An exploratory permutation test found no evidence that the paired differences depended on presentation order ($p=0.388$).

Participants' qualitative feedback did not suggest that either visualization was uniformly easier to use. Instead, usability depended on whether participants could map the available controls and visual encodings onto a recognizable analytical question. \toolname{} showed more specialized information, so it required some time to learn at first. However, filtering reduced this complexity: P6 described the character filter as a ``\textit{simplifying mechanism},'' and P3 found it easier once she limited the display to one character or focalization category. The Story Ribbons visualization presented a different usability problem. Although some participants perceived its visual form as cleaner, they were often less certain about how its encoding related to the underlying text or what analytical action they were expected to take. P10 commented:

\begin{quote}
``It doesn't feel user-friendly. [\ldots{}] I feel like I'm not able to directly connect this with the text because I'm not able to see these insights with alongside text.'' (P10)
\end{quote}

A common usability concern across the two tools was the reliability of annotations and explanations. Six participants (P2, P4, P6--7, and P9--10) questioned the accuracy or interpretation of LLM-generated annotations and explanations. Despite these concerns, eleven participants (P1--P6, P8--9, and P11--13) stated that the explanations were helpful for understanding and critically assessing the focalization annotations.

\subsubsection{RQ3: Cognitive load}
Workload was similar in both conditions (\autoref{fig:sus-and-nasa}). \toolname{} received a mean NASA-TLX score of $M=48.91$ ($SD=13.79$, 95\% CI $[41.73,\allowbreak 56.35]$), compared with $M=49.04$ ($SD=14.65$, 95\% CI $[41.22,\allowbreak 56.28]$) for Story Ribbons. The difference was not statistically significant (Wilcoxon signed-rank test, $W=42.0$, $p=0.827$). An exploratory permutation test found no evidence that the paired differences depended on presentation order ($p=0.988$).

Participants described two sources of cognitive load: the amount of information on screen and the interpretive concepts behind it. With \toolname{}, participants first had to keep characters, focalization types, facets, and narrative granularity in mind at once. P1 described this directly about FocalLens:

\begin{quote}
``this is a very cool tool, but I am finding it, like, a bit intimidating to use, or [\ldots{}] there are, like, so many different things to search for that it feels overwhelming just to start, I guess.'' (P1)
\end{quote}

Beyond the amount of information, participants also had to learn the conceptual distinctions represented in the system. While internal and external focalization were often understood relatively quickly, participants frequently found the boundaries among perceptual, ideological, and psychological focalization less clear. This conceptual effort was compounded when many characters and categories were displayed simultaneously. 

The Story Ribbons visualization shifted the cognitive burden from learning focalization terminology toward understanding a larger set of heterogeneous metrics and encodings. Participants repeatedly asked how variables such as importance, conflict, sentiment, color, and network geometry were derived. For instance, P2 asked how ``\textit{conflict, importance, and sentiment}'' were calculated, while P11 distinguished relatively interpretable measures from the less transparent notion of importance:

\begin{quote}
    ``I don't see enough information about the importance. Like, I kind of understand conflict, emotions, um, and length, like, that's very clear, but for importance, as a literary scholar, I'm like, wait, what? How?'' (P11)
\end{quote}

Similarly, P10 remarked that ``\textit{Importance is such a subjective meter,}'' while P7 questioned a low importance score for the scene in which Nick first meets Gatsby, arguing that the meeting is ``\textit{pretty [\ldots{}] pivotal}'' to the rest of the story. 

\subsubsection{Case Studies for Creative Writing}
\fix{During the user study, writers who shared their own drafts explored them using FocalLens towards the end of the session. This was on top of the analysis they already performed using the two tools. Below, we demonstrate usage scenarios for creative writing using some concrete examples from the study.} 

\begin{figure}[!ht]
    \centering
    \includegraphics[width=\linewidth]{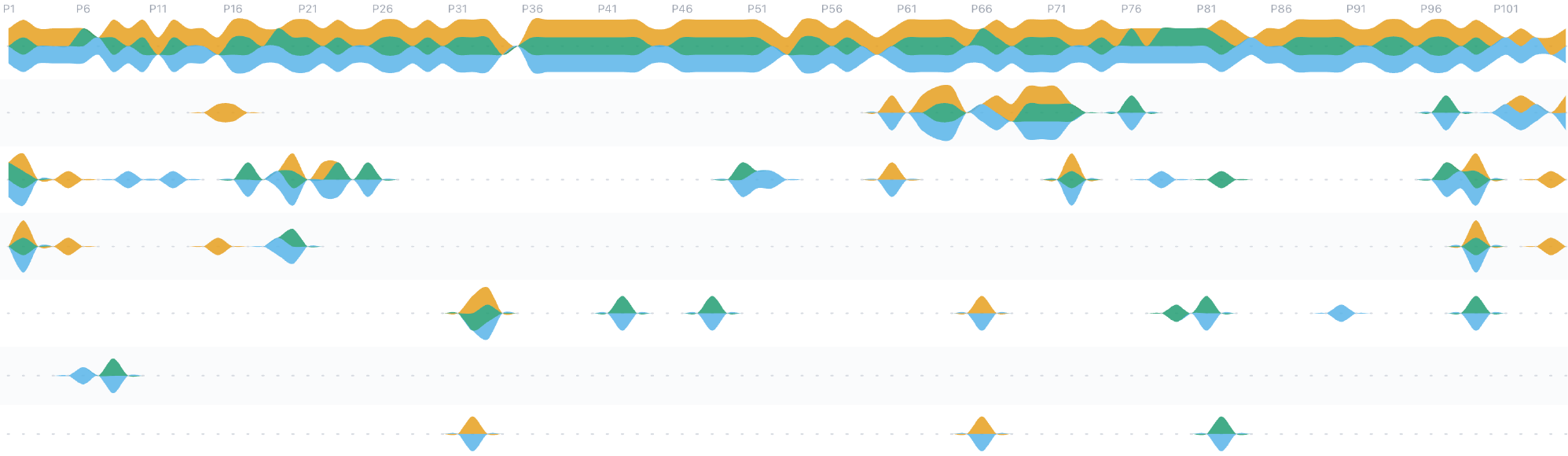}
    \caption{\textbf{Visualization of P3's close-third-person narrative.} The main character (top row), an animal portrayed through close-third-person narration, appears most frequently and exhibits multiple focalization categories. Character names and story text are omitted to protect the participant's original work.}
    \label{fig:creative-writing-case}
\end{figure}

P2 appreciated being able to view ``the data of [their] manuscript'' rather than repeatedly navigating between pages. P2 suggested that the visualization could help identify whether a minor character received sufficient interiority, asking, \textit{``am I forgetting to, like, give a character interiority in this story, like, a minor character, and then you could see it?''} Similarly, P3 explored a draft written in close third person, stating, \textit{``The story is in close third, so I want to see if I was faithful to that.''} P3 used the linked visualization (\autoref{fig:creative-writing-case}) and source text to examine whether the displayed focalization patterns aligned with this intention.

P6 explored her own short work of creative nonfiction and praised the relationship graph (similar to~\autoref{fig:relationship-graph}), describing it as ``a really simple and clear visualization of character depth.'' She explained that the graph helped distinguish ``characters that we are just watching do things, as opposed to who are the characters that are having insights, who are the characters making judgments.'' She also appreciated how it allowed her to examine characters along a progression ``from perhaps more simple or straightforward to more complicated.'' However, she also questioned how the annotations reflected her narrative choices. She noticed that most of her story was labeled as external focalization, and that the use of the first person seemed to be the only cue for internal focalization; yet in her genre, ``I'' does not necessarily indicate internal focalization. Her exploration illustrates how writers can use FocalLens both to reflect on how focalization contributes to characterization and to critically examine whether the displayed annotations align with their understanding of their writing.

\subsubsection{Suggested extensions}
Participants suggested additional visualization and navigation mechanisms to support literary analysis and writing-related exploration. P5 proposed cross-filtering focalization with other narrative dimensions, such as location and character interactions, to examine how a character was focalized when appearing in particular locations or interacting with other characters. Extending this idea of comparative exploration across different narratives, P11 suggested a dedicated view for comparing characters from multiple stories:

\begin{quote}
    ``if there is a comparison between just the main character, one main character of each story, that would be helpful, because that allows me to see the different styles specifically.'' (P11)
\end{quote}

Such a view could support comparative literary analysis by allowing scholars to examine focalization patterns across characters, works, or authors without repeatedly switching between narratives. Participants also emphasized the need for fluid movement between overview and close reading. P3 preferred chapter-level views for longer narratives and wanted to select a region of the visualization and immediately inspect the corresponding source passage. Together, these suggestions point toward a multi-scale design that preserves FocalLens's overview of perspective while supporting richer plot context, clearer event structure, and seamless drill-down to chapters, scenes, and source passages.

%% file: table/table2.tex
\begin{table*}[t]
\centering
\scriptsize
\setlength{\tabcolsep}{4pt}
\caption{Summary of insight types, definitions, examples, and counts. A single insight can have multiple insight labels. The multiple labels are appended at the end of each example insight. Bars show counts for \swatch{barA}~FocalLens and \swatch{orange}~StoryRibbons.}
\label{tab:types}
\begin{tabularx}{\textwidth}{>{\raggedright\arraybackslash}p{1.5cm}
                             >{\raggedright\arraybackslash}p{3.3cm}
                             >{\raggedright\arraybackslash}X
                             l}
\toprule
\textbf{Type} & \textbf{Definition} & \textbf{Examples} & \textbf{Counts} \\
\midrule

    I1. Overall Plot Structure & Insights about repeated patterns or progression through multiple parts of the story. & 
    \exrows{Um, so \textbf{these wave patterns} are demonstrating, like, what are the points at which, like, the emotions are high, [\ldots{}] the focalization, like, there's \textbf{some event that's happening to trigger}. (P12) (also in I2)}{Frankenstein is a story in which [\ldots{}] they're \textbf{constantly moving around}, which makes sense. It's a [\ldots{}] \textbf{chase story}, in a way. (P6) (also in I4)} &
    \counts{12}{22} \\
    \midrule

    I2. Story Perspective & Insights about perspectives, focalization, and narration. & \exrows{[\ldots{}] there are other characters whose \textbf{focalization} is really important, and [\ldots{}] this tool reflects really \textbf{beautifully} is the way [\ldots{}] the chapter sort of \textbf{looks through Gertie McDowell's eyes} until halfway through the chapter, it flips over to looking through \textbf{Leopold Bloom's eyes}. (P4) (also in I6)}{Victor's presence at the scene \textbf{to hear this story from the creature} is sort of crucial. [\ldots{}] in fact, Victor is the \textbf{one who's recounting the story to Robert Walton}. (P5) (also in I4)} &
    \counts{38}{9} \\
    \midrule

    I3. Scene Dynamics & Insights about a single scene or paragraph to explain how it unfolds, such as how a pause, a reaction, or a change in pace builds tension or marks a turning point. & \exrows{\textbf{This is a very particular scene. In here} where one of the main points of conflict or shock happens [\ldots{}] a father sexually assaults his daughter. And you only \textbf{know it happens, but when you see it in action here}, it is striking. (P10) (also in I2 and I5)}{I'd maybe be like, what are the \textbf{three most important scenes} for \textbf{this certain character}, and then I would pick those. Like, I read something like, oh, yeah, well, like, \textbf{Miss Nash was there}. And then I was like what \textbf{other scene she was?} Oh, yeah, at the Woodhouse, you know, home, like, there we go. (P9) (also in I1)} &
    \counts{7}{3} \\
    \midrule

    I4. Character Interaction & Insights about characters appearing together or not, and how their interactions create climax or anticlimax. & \exrows{[\ldots{}] So this is all of \textbf{her different relationships with different people}. Maybe, like, \textbf{Emma's relationship with herself} is, like, a very interesting [\ldots{}] kind of point to think about. [\ldots{}] So \textbf{I'm thinking about the narrator's relationship with Emma}. (P12)}{I like  \textbf{the character web} of that. Because again, \textbf{you can see like who's interacting with whom}. And. I think that would help like showcase that, for example, here, \textbf{some characters do not interact with each other at all}. (P2)} &
    \counts{4}{12} \\
    \midrule

    I5. Character Traits and Emotions & Insights about characters' importance, emotions, sentiments, demographics, social biases, look and feel. & \exrows{\textbf{It helps externalize certain things}. [\ldots{}] \textbf{we have a consistent first person narrator in this text}. And so \textbf{there's always going to be a stream of consciousness and an internal focalization happening anyway}. And then they can [\ldots{}] write a paper on one or two modes of, like, \textbf{external censorship or external}, [\ldots{}] validation or not validation that \textbf{she's getting from her husband}, \textbf{her relationship to her children, the woman in the wall, whether that's internal or external}, all of these things [\ldots{}] just open up these avenues for more research, having it kind of laid out in this way. (P12) (also in I2)}{\textbf{She basically became crazy to liberate herself from the world}, right? \textbf{She enters the wallpaper, she becomes free. And that's the only exit from the thing}, right? So it's like, in some ways, \textbf{psychosis was the best thing for her.} (P9) (also in I1)} &
    \counts{5}{14} \\
    \midrule

    I6. Writing Style & Insights about the author's overall writing style, such as word choice, tone, and techniques like free indirect speech. & \exrows{what I'm doing right now is, like, I want to see if \textbf{Emma in Emma [\ldots{}] her internal focalization is much more evident than in other stories}, which is why I'm kind of trying to select that. [\ldots{}] why we say \textbf{Jane Austen has this style.} Quantitatively, uh, because she has, like, \textbf{more [\ldots{}] internal focalization than others}. (P11) (also in I2)}{N/A} &
    \counts{9}{0} \\
    \midrule

    % I7. Dialogue and Narrative Discourse & Insights about the style of dialogue and discourse, and how that impacts the story and narrative. & \exrows{Although, in fact, here he actually sort of is present as a focalizer, in the sense that he's the one who recounts what the creature has said, and so we can only rely on his word for it. (P4)}{But the whole novel is in some way written down by Colonel Walton on the ship, and he returns at the very end. So the whole novel is actually sort of in quotation marks, being told by Colonel Walton. (P4)} &
    % \counts{7}{6} \\
    % \midrule

    I7. Personification of Objects & Insights about objects, places, animals, or abstract ideas that are given human qualities, emotions, or agency, and what role they play in the story. & \exrows{N/A}{the \textbf{billboard} is not really what the novel is interested in. It's interested in the \textbf{symbolism of these gigantic, watchful eyes}, you know, looming over. (P4)} &
    \counts{0}{3} \\
\bottomrule
\end{tabularx}
\end{table*}

%% file: sections/07-discussion.tex
\section{Discussion}
\label{sec:discussion}
We discuss design implications learned from this work and future directions below.

\subsection{Implications for Visualizing Narratives}
Our findings indicate that \toolname{} and Story Ribbons facilitate complementary insights. Thus, a natural next step is to develop a way to merge them together. However, this is not straightforward, considering the limited visual channels and the need to reduce complexity. One of our initial design goals (DG2) was to adopt storyline visualization, the primary visual representation of Story Ribbons. While \toolname{} resembles storyline visualization, we had to abandon character grouping in favor of encoding focalization categories to reduce the complexity of the tool. This design choice was reflected in the user study results.

One possible approach is to incorporate the tools as separate modules into a single tool or to provide choices for the representation~\cite{hoque2023portrayal}. However, this design might become too open-ended and overwhelming for users. It also may not scale well to more than two tools. Another possible approach is to follow a hypothesis-based approach. Users will specify the type of analysis they want to conduct or the questions they want to answer, and the system will facilitate that visual and textual evidence. There are opportunities to develop formal notations~\cite{DBLP:journals/tvcg/MoritzWNLSHH19} or declarative grammars~\cite{DBLP:journals/tvcg/ParkDFE18, DBLP:journals/tvcg/SatyanarayanMWH17} to facilitate such a hypothesis-based approach.

\subsection{Implications for Writing Support Tools}
Generative AI has reshaped writing support tools. However, there are concerns among writers that these techniques are harmful to writers and readers, and, above all, to good literary work. \fix{During the 2023 Writers Guild of America (WGA) strike, Hollywood writers demanded contractual protections against being replaced by AI~\cite{wga2023agreement}}. The main reasons for these concerns are the human-like text-generation capabilities of LLMs. However, we believe this is not the only application of LLMs in literary work. Rather than positioning LLMs as co-authors, \toolname{} positions itself as an analytical instrument for uncovering narrative structures. Thus, this project departs from existing research on writing and creativity support tools. As a result, in the user study, we did not receive concerns about ownership, agency, and copyright, which are common in LLM-based writing support tools~\cite{DBLP:conf/chi/HoqueMGSCKE24, DBLP:conf/ACMdis/GeroLSD26, DBLP:journals/tochi/DraxlerWLHSBW24}. While the results are still preliminary, we hope that our work will motivate future research to explore the role of LLMs as analytical agents rather than generative agents in supporting creative writing. 

\subsection{Implications for NLP Research}

To our knowledge, no established benchmark exists for identifying focalization in narrative texts, apart from an initial effort by Hicke et al.~\cite{hicke2025says}. In our small-scale experiment, GPT-5.6 reached a macro-F1 of 0.831 on our reference set, but its labels diverged from the reference annotations in some cases. Participants in the user study likewise identified a few instances in which they disagreed with the model's predictions. They also identified a few explanations as misleading or incorrect. This points to the need for a thorough investigation of the model performance and error analysis. An important consideration is that focalization is an interpretive task for which human annotators may also disagree. Accordingly, model performance should be contextualized relative to human annotator agreement rather than evaluated solely by classification accuracy. Comparing model--reference and human--human agreement using consistent metrics, such as Jaccard similarity and Cohen's $\kappa$ where applicable, could help establish whether model predictions approach the agreement observed among human annotators. Such comparisons would provide a more meaningful account of model performance on interpretive tasks. Future work should expand the reference dataset across literary genres, writing styles, and narrative structures, with multiple independent annotations to characterize interpretive disagreement. Further error analysis could identify cases where models struggle with ambiguous focalization, nested narratives, or distinctions between narrator and focalizer. Beyond focalization, future research can focus on creating benchmarks for a wide variety of narrative analysis tasks.

%% file: sections/08-conclusion.tex
\section{Conclusion}
\label{sec:conclusion}
We presented \toolname{}, a visual interface that leverages the concept of focalization to support narrative analysis. Grounded in a formative study with literary scholars, \toolname{} links focalization patterns across characters and narrative units to the source text with a stream-graph-like representation, supported by an LLM. Our study shows that \toolname{} facilitated insights about story perspectives and overall writing style, complementing Story Ribbons, which better reveals plot structure and character interaction. As a technology probe, \toolname{} is an initial step toward visualizing the complex narrative components that existing tools leave implicit.